\documentclass{workshop}
\usepackage{natbib_mod}
\usepackage{amssymb}
\usepackage{amsmath}
\def\apj{ApJ}
\def\apjl{ApJL}
\def\apjs{ApJS}
\def\aap{A\&A}
\def\nat{Nature} 
\def\mnras{MNRAS}
\def\aapr{A\&ARv}
\def\araa{ARA\&A}
\def\pasj{PASJ}

\begin{document}


\title{Recent Constraints on Jet physics and Properties Obtained from 
High Energy Observations of Microquasars} 

\author{
J\'er\^ome Rodriguez$^1$ 
\\[12pt]  
%
$^1$  Lab AIM - CEA/IRFU-CNRS/INSU-Universit\'e Paris Diderot, 
CEA DRF/IRFU/SAp,  F-91191 Gif-sur-Yvette, France.\\
%
\textit{E-mail: jrodriguez@cea.fr} 
}

\abst{The recent detections of microquasars at energies above a few hundred keV up to the TeV in one case has stimulated a strong interest 
 and raised several questions.  
How are the MeV, GeV, and even  TeV emissions, produced? What are the emission processes and in-fine 
the physics and media at the origin of the broad band spectra?  What is the content of these media, and how are they powered? 
These sources are machines accelerating particles and matter to very high speed in collimated jets. While these are now known for more 
than twenty years (mainly with radio observations) and their behavior known to be tied to the accretion processes 
onto the compact object (studied via X-ray observations), their potential influence at high energy is just being recognized. 
In this review I will present a selection of recent results obtained with XMM-Newton, INTEGRAL, Fermi on 4U 1630$-$40,  
Cygnus X-1, and V404 Cygni and will discuss the possible interpretations of these results. }

\kword{accretion, accretion disks --- jets sources --- black hole physics --- X-rays: binaries --- Gamma-rays: observations --- stars: individual: Cyg X-1, V404 Cyg, 4U 1630-47 }

\maketitle
\thispagestyle{empty}
\section{Introduction}
Microquasars (MQs) are Galactic X-ray binaries showing episodes of ejections in certain phases
of their activity. Most MQs host a black hole as compact object, and 
the majority are transient sources that spend most of their lives in a dormant, so-called quiescent state. 
They are detected (usually in X-rays) when they enter into month to year long periods of activity called 
outbursts. All MQs are multi-wavelength emitters. The radio to infrared (IR) emission is attributed to relativistic jets, 
either a persistent compact jet or in form of large scale discrete ejections. The X-rays probe the inner accretion flow(s):  
the accretion disc ($\sim$0.1--10~keV) and the so-called ``corona'' (10--200 keV)  \citep[e.g.][]{Fender06b,Remillard06}. 
The relative contributions of these two media seen in the X-ray spectra have led to the classical classification
into spectral states. The two canonical ones are the ``(high) soft state'' (HSS) and the ``(low) hard state'' (LHS). In the former the 
X-ray spectrum is  dominated by the thermal emission from the disc, and in the latter it is dominated by a (cut-off) power 
law with a hard ($\Gamma$$<$2) photon index attributed to inverse Comptonization (IC) of the soft disc photons by  the 
coronal electrons. The past 25 years have allowed us to enrich this initial vision, and, in particular, to find 
strong connections between X-rays (accretion) and radio-IR (ejection) \citep[e.g.][to cite but a few]{mirabel98, fender98,Corbel00, fender01, Corbel03, rodrigue03_1550, rodrigue08_1915b, Corbel13b}: a powerful compact jet is always present in the LHS, while 
transitions from the LHS to the HSS are accompanied by discrete ejections.\\
 \begin{figure}[ht]
\includegraphics[width=6cm]{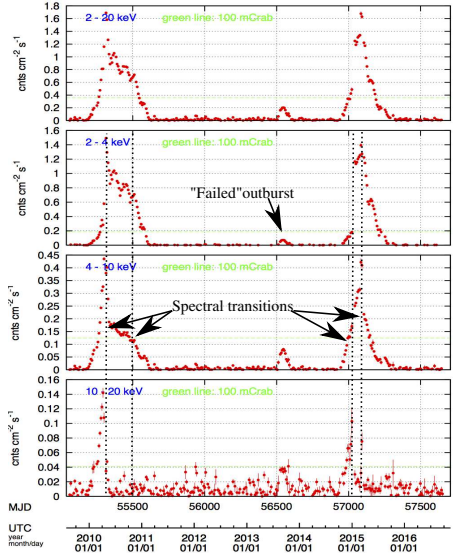}
\caption{MAXI light curves of GX 339$-$4 obtained from the  MAXI online data at http://maxi.riken.jp/top/ }
\label{fig:LCs}
\end{figure}
\indent The picture may have recently been completed by the detection of  a few sources at 
energies above 1~MeV with  OSSE and INTEGRAL, and/or in the GeV domain with Fermi and Agile, 
and possibly in the TeV range with the Magic telescope. 
These detections pose questions regarding the media and physical processes at work, with answers 
that will certainly impact our understanding of particles acceleration,  particle-matter and particle-particle interaction, 
and eventually will help us constrain feedback processes in the interstellar medium. \\
\indent As mentioned above these sources are either transients, and they are anyway highly variable. 
One thus needs alerts from surveying instruments in order to be able to conduct multi-wavelength, multi-instrumental 
campaigns during the most relevant periods of their activity.  Before presenting recent high energy observations and analysis of 
a few MQs, I will first start this review by presenting the obvious but necessary needs of all sky surveying/monitoring  
instruments without which most of the results of the past 30 years will not have been obtained. 

\section{Transient and variable sources need to be surveyed: the central role of all-sky monitors} 
\label{sec:asm}
\begin{figure*}[!t]
\centering
\includegraphics[width=16cm, height=7cm]{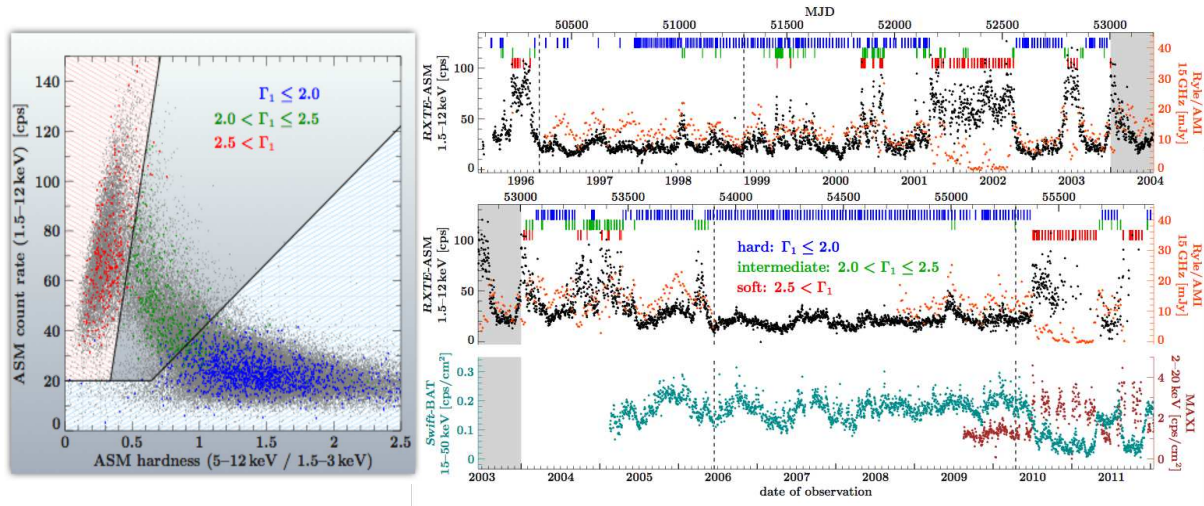}
\caption{{\em{Left:}} Flux-Hardness diagram based on the RXTE/ASM data. The different regions show different spectral 
states as confirmed with a spectral analysis.  
{\em{Right:}} RXTE$\slash$ASM, MAXI, and Ryle/AMI 15 GHz light curves of Cygnus X-1 over 16 years. Adapted from \citet{grinberg2014}.}
\label{fig:Hardness}
\end{figure*}

MQs  have unpredictable behavior. They enter into outburst and/or show (spectral) transitions at ``random" moments. 
These are ideal times to probe the interactions and causal relations between 
the different media (jet-disc-corona-companion). It is thus necessary to keep an eye on the entire sky to detect 
new or know object as early as possible during their outbursts, or simply check for any spectral transition  
that may 'act' as a smoking gun for major change of configuration of the inner flows and possible ejection events. Wide field monitoring instruments (hereafter ASM for all sky 
monitors of any types) are the only instruments to do so.\\
\indent Fig.~\ref{fig:LCs} shows the MAXI light curves covering a multi-year survey of GX 339$-$4 that, in 
many ways, can be considered as the prototype MQ \citep[e.g.][]{corbel13}. 
Over the 7 years of X-ray coverage, one can clearly 
distinguish three outbursts separated by long intervals of quiescence. The outbursts, 
moreover, show very different profiles, with two long ($>$6 months), bright ($>$100 mCrab at all energies) 
episodes, and a shorter and much fainter one (dubbed ``failed" in Fig~\ref{fig:LCs}). The spectral capabilities 
of ASMs permit to have an overview of the spectral dependence
of the sources light curves, build hardness ratios and hardness intensity diagrams (HID). This allows to survey 
the sources spectral evolution/transitions and classify their states (Fig~\ref{fig:Hardness}) to eventually 
trigger/organise observing campaigns during specific periods of activity. \\

\indent It is thus  possible to use the entire database of any ASM to classify 
 a source's behavior from multi-color light-curves and HID without a full model dependent approach to the data 
 \citep[e.g][in the respective cases of GRS 1915+105 and Cyg X-1; see Fig.~\ref{fig:Hardness}]{punsly13a,grinberg2014}. 
 This also allows to fill the temporal gaps and/or uneven 
 temporal sampling of dedicated pointed observing campaign made with more efficient,  smaller field instruments (XMM-Newton, 
Chandra, RXTE, Nustar, for instance). This is an approach that we followed and that allowed us to refine the study of the high energy 
properties of Cyg X-1 \citep{rodrigue15_cyg}. 

\section{Cyg X-1 and the possible origin hard spectral tails seen in (some) microquasars}
\label{sec:cygx1}
 While emission above $100$~keV from microquasars  is known for more than 20 years \citep[e.g.][to cite just a few]{laurent93,churazov94,grove98,mcconnell00}, 
 no consensus on its origin has been reached yet. It can be  interpreted as 
 due to IC, either purely thermal \citep[spectrum with an exponential cut-off at about 100 keV,][]{titarchuk94,rodrigue03_1550, marion1753}, 
 or thermal/non-thermal in an hybrid plasma \citep[e.g][]{zdziarski01,delsanto13}, or from thermal IC from two different populations of 
 electrons  \citep[e.g.][to fit the 1E 1740.7$-$2942 INTEGRAL spectrum]{bouchet09}. 
 In the LHS, jet emission can also be invoked \citep[e.g][]{markoff05} 
 through various radiation processes (Comptonisation, Synchrotron and Synchrotron-Self Compton - SSC). Apart from 
 pure thermal IC,  all these models predict similar 20--1000 keV spectral shape, and  a single 
 spectral analysis, even with the high quality data obtained today, does not permit to break the model 
 degeneracy \citep[see also][]{Nowak11}.\\
\indent Using the large Cyg X-1 INTEGRAL archive \citet{laurent11} with IBIS, and  \citet{jourdain12} with SPI ,
reported the detection of a high energy tail extending to about 1 MeV in addition to a standard IC component below 
typically 200--300 keV, and a strong polarized signal above $\sim$400~keV.
The large polarisation fraction reported ($\gtrsim60\%$) led both teams to conclude that the origin of this 
$>400$~keV tail was Synchrotron emission coming from a compact jet \citep{laurent11, jourdain12}.\\
\indent In both studies, however, the entire INTEGRAL data acquired before 2009 was used, mixing thus all spectral 
states. As no radio data was considered by any of the two teams the presence of a jet could not be proved either 
although Cyg X-1 is a known jet source. To go further, and discriminate the potential state dependence of the results, 
we classified the whole INTEGRAL and AMI-Ryle databases 
into three basic states based on  the ASM-based model independent classification 
\citep[][in Fig.~\ref{fig:Hardness}, Sec. \ref{sec:asm}]{grinberg2014}. Stacked high energy spectra, polarigrams and radio 
properties were then obtained and studied for all three states. 
\begin{figure}[!t]
\centering
\includegraphics[width=6cm]{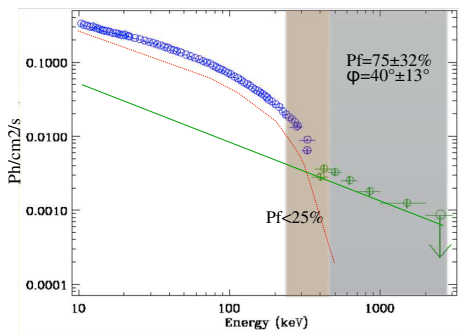}
\caption{LHS INTEGRAL spectrum of Cyg X-1. The shaded parts indicate the spectral range were 
polarization measures were obtained, with the levels obtained in each cases.  The red (resp. green) line represents the 
IC (resp. power law) component. Adapted from \citet{rodrigue15_cyg}.}
\label{fig:cyg}
\end{figure}
 A high energy tail, dominating the spectrum above $\sim$400 keV,  is visible only in the LHS 
 \citep[Fig.~\ref{fig:cyg}][]{rodrigue15_cyg}.  A high degree ($\sim75\%$) of polarisation is also measured in this state 
 at energies where the tail dominates, while only an upper limit is obtained at lower energies, where the spectrum is 
 dominated by IC, and in the other states. The radio observations independently indicate a high level of emission (a mean 
 of about 13 mJy) in the LHS while in the HSS we report a mean level of 3 mJy, probably due to residual emission from discrete 
 ejections occurring at LHS to HSS transitions. Our refined analysis therefore clearly confirms the presence of both a highly polarized 
 $>$400 keV tail and a compact radio jet during the LHS. We interpret this tail as  Synchrotron emission 
 from the jet  \citep{laurent11, jourdain12,rodrigue15_cyg}. \\
 \indent Alternative possibilities, such as hybrid thermal-non thermal 
 Comptonization in the corona \citep{zdziarski14a}, 
 or Synchrotron from the corona \citep[see][for all details]{romero14}, have, however, also been suggested. The clear  
 detection of Cyg X-1 with Fermi/LAT up to $\sim$10  GeV \citep{bodaghee13,zanin16,zdziarski16_cygx1} in the LHS, in addition to the 
 possible TeV detection with MAGIC during a flare \citep{albert07}  renders the non-jet interpretation less likely \citep{zdziarski14a, pepe15}. 
 In particular, and as I discuss in Sec.  \ref{sec:model},  the coronal model implies a configuration that is not entirely clear, and  
  the MeV-GeV emission is better modeled by emission from a jet \citep[][Sec.~\ref{sec:model}]{zdziarski14a, pepe15,zanin16,zdziarski16_cygx1, zhang17}, than by a model invoking IC up to the MeV. \\
 
 \begin{figure}[th]
\centering
\includegraphics[width=7cm]{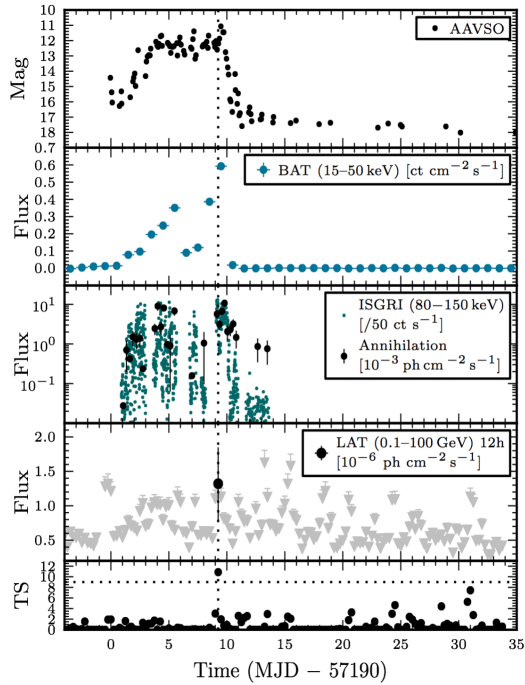}
\caption{Multiwavelength light-curve of V404 Cyg during the June 2015 outburst. From top to bottom AAVSO, Swift BAT, 
INTEGRAL/ISGRI and INTEGRAL SPI 511 keV, Fermi/LAT light curves, and statistical test of the Fermi data. 
 Adapted from \citet{loh16,siegert16}, courtesy A. Loh.}
\label{fig:v404}
\end{figure}

\section{The extreme case of V404 Cygni}
\label{sec:v404}
The low mass microquasar V404 Cygni entered into  outburst on June 15, 2015 after more than 25 years of quiescence. It was detected 
with all X-ray wide field instruments \citep{Barthelmy15, Negoro_Atel7646,Kuulkers_Atel7647}. It was soon recognized that the source 
had a dramatic behavior, showing intense flares (up to 50 Crab in the 20--40 keV range) on timescales of about an hour \citep[e.g.][Fig. ~\ref{fig:v404}]{rodrigue15_V404,roques15,jourdain17}. It was also clearly detected up to high energy early in the outburst \citep{roques15_atel}, and 
a multi-instrumental follow-up and analysis showed probable detection of the source with Fermi in the GeV range 
\citep[Fig. ~\ref{fig:v404}]{loh16}.\\
\indent Until then all GeV binaries were high mass systems with either a Be or a Supergiant companion. In these systems, the GeV 
emission is explained by IC scattering of the stellar (UV) photons on relativistic electrons near the compact object  
\citep[this model is known as the leptonic model, e.g][and references therein, Sec. \ref{sec:model}]{bednarek13,dubus13}. In V404 Cyg, 
however,  the secondary is a $0.7_{-0.2}^{+0.3}$\,M$_\odot$ K3 III companion \citep[e.g.]{Khargharia10}. 
In the context of the leptonic model, one cannot expect GeV emission from this source. The discovery of a variable transient feature
in the 511 keV region, attributed to $e^-/e^+$ annihilation further brings the question about the production of pair plasma \citep{siegert16}, 
and the potential role of a jet (that is detected at other wavelengths). 
The annihilation flux seems correlated with the X-ray flaring activity, which itself  somehow correlates with the flares 
seen at longer wavelengths \citep{rodrigue15_V404, siegert16,loh16,munoz16}. 
\citet{loh16} report a $\sim$4$\sigma$ detection in the 0.1--100 GeV Fermi observations at a position coincident with V 404 Cyg.
The detected emission at GeV energies occurs in conjunction with the brightest hard X-ray flare, itself associated 
with a spectral change in the source. Both events appear soon after ($\lesssim$6 hr) a giant  radio flare was detected \citep{loh16}. 
The temporal and spatial 
coincidence of all events strongly support a $\gamma$-ray flare from V404 Cyg during a major flare.  The simultaneous detection 
of the 511 keV annihilation line and $\gamma$-rays of higher energies imply that the latter 
emission occurs far from the corona,  pointing in the direction of jet emission \citep{loh16}. The multi-wavelengths properties 
near the Fermi detection are actually compatible with a Blandford-Znajek (BZ) powered jets interpretation during states when the 
inner accretion disk is magnetically arrested \citep{loh16}.

\section{Models for high energy emission and jet contents}
\label{sec:model}

The origin of the high energy ($>$MeV) emission is debated. Two main families of models co-exist and  
predict strong differences at very high energies only (TeV),  a domain were there is, to date, only one marginally significant 
detection of Cyg X-1 \citep{albert07}. In the first family (leptonic models) the high energy emission is due to 
IC scattering of the strong stellar photon field (until 2015 all GeV detected binaries were high mass systems) on 
relativistic electrons. The second family (hadronic models) invoke the presence of a jet, and the MeV-GeV is 
instead due to post break Synchrotron and SSC radiations \citep[see e.g.][and references therein]{pepe15}. 
Leptonic models are popular in the context of $\gamma$-ray binaries where the companion is a high mass star and 
the primary  a pulsar. \\
\begin{figure}[htbp]
\includegraphics[width=8cm]{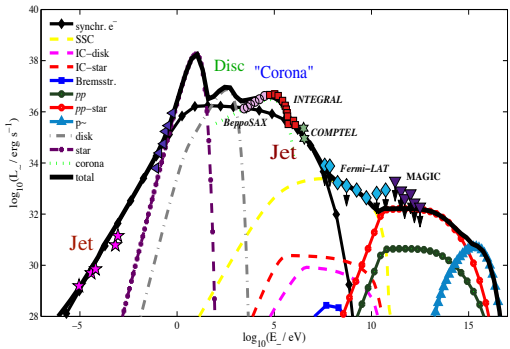}
\caption{Lepto-hadronic modeling of the Cyg X-1 broad band spectrum with a set of parameters favoring Synchrotron emission and 
SSC from the jet as the origin of the MeV-GeV component.  Adapted from \citet{pepe15}.}
\label{fig:cygmod}
\end{figure}
In the LHS of MQs the radio-IR emission (Fig.~\ref{fig:cygmod}) is well explained by  self-absorbed Synchrotron 
radiation from a population of electrons in a compact jet \citep[e.g.][]{Blandford79,fender01,Fender06b,Corbel13b}. Here the presence of the relativistically accelerated particles in a strongly 
collimated outflow might change our interpretation of the origin the high energy emission(s). 
The strong polarization fraction detected in the $\gamma$-ray tail of Cyg X-1 excludes IC \citep{rodrigue15_cyg}, but not necessarily 
coronal emission \citep{romero14}. However the latter explanation contains intrinsic difficulties that led us to favor the jet 
emission. The high level of polarization measured with INTEGRAL needs an highly ordered magnetic field which is likely to be found 
close to the base of the jet, where high energy emission could also occur\footnote{It is interesting to note that \citet{zhang17}
could also reproduce the polarized hard tail with a jet model involving both large and small scale turbulent magnetic fields. }. 
Instead, Synchrotron photons produced by coronal electrons
would either have to undergo many Compton scattering if produced deep into the corona, and the polarisation would be lost, or 
they would be produced too far from the highly ordered field to explain the high degree of polarization we measured.  
\\
\indent \citet{pepe15} recently applied a lepto-hadronic model to represent the broad band spectrum of 
Cyg X-1, the  hadronic component being 
required by the TeV flare detected by MAGIC. While globally, the two sets of parameters they use fit the data rather well, 
the one based on coronal IC and IC of the companion star fails to represent the MeV tail, while this component 
is well fitted when considering direct jet Synchrotron and SSC instead (Fig.~\ref{fig:cygmod}). The recent detection 
of MeV and GeV emission during the V404 Cyg outburst \citep{siegert16, loh16, jourdain17} implies similar conclusions, especially 
in a system where the secondary has no wind and a much weaker and cooler photon field than high mass systems.  \\
\indent Knowing the composition of jets is of prime importance as it permits 
to have a proper estimate of their total kinetic power and in-fine their feedback (matter and energy) into the ISM. 
 Baryonic jets should indeed cary significantly more energy than pure leptonic ones, and this of course puts even 
 stronger constraints on the mechanisms responsible for their production and acceleration. 
It seems widely admitted that jet might cary a very significant part of the accretion energy \citep{fender01c, Gallo05}, in 
the LHS. Jet acceleration is thought to either be based on the extraction of the BH spin (BZ) or magnetic energy from the 
disc (Blandord \& Payne mechanism). Until recently, SS 433 was the only MQ were baryonic jets had been 
detected \citep[e.g.][]{kotani94}, but this source 
is a peculiar MQ in many other ways, and this could be just another of its peculiarities.  In this respect the recent  XMM-Newton  detection 
of Doppler shifted emission lines in 4U 1630$-$47~  \citep{diaz13}, a more typical MQ,  may shed new light on this aspect.
\citet{diaz13} could use a simultaneous coverage at radio wavelengths that showed that the Doppler shifted 
lines are coincident with the detection of a jet. The spectral analysis of the X-ray lines are compatible with being 
emitted from a $\sim$0.7 $c$ outflow. The flux ratio between the red and the blue shifted lines is, furthermore, consistent with 
Doppler boosting \citep{diaz13}. This exciting result may imply that typical MQ jets do carry baryons, and that, therefore, their 
energetic budget is even more important than previously thought. Additionally, and as discussed by \citet{diaz13}, the detection 
of a baryonic jet favours models of accretion disk powered jet rather than BZ jets. A baryonic
jet implies that 4U 1630$-$47 should be a strong source of $\gamma$-rays, and this source is not a known as an high energy emitter. 
These lines have, furthermore never been confirmed  since their first report with new Chandra or 
re-analysis of the XMM-Newton observation \citep{neilsen14,wang16}. It is therefore not possible to fully conclude on this issue, 
which leaves the mystery complete. 

\section{Conclusions}
Jets are clearly important channels for energetic, material redistribution and feedback into the ISM. Their 
detection at high energy will certainly modify our understanding of MQs in general. 
The chosen results described in this short review are just the tip of the iceberg, and certainly do not allow to have firm 
conclusions on most of the questions I raised in the introduction. Soon, the advent the CTA at TeV energy, and/or the detection 
of neutrinos from a MQ,  in conjunction with observations at all wavelengths may help to make 
a step forward into many of these aspects. One should still keep in mind that studying  the accretion-ejection physics relies 
on multi-instrumental  campaigns triggered at appropriate times. This cannot be done without constant survey 
of the transient sky, and our ability to understand better these objects definitely rely on all sky monitorings such as those made  
with RXTE/ASM in the past, or currently with MAXI, and Swift.  

\section*{Acknowledgements}
I Warmly thank C. Gouiff\`es, P. Laurent, \& A. Loh for useful discussions and comments on this manuscript. Partial funding support is provided through the CHAOS project (ANR-12-BS05-0009), and  UnivEarthS Labex  (ANR-10-LABX-0023 and ANR-11-IDEX-0005-02), and by the French Space Agency (CNES) in support to the analysis of the INTEGRAL data. 

\label{last}

\end{document}